\documentclass[fleqn,usenatbib,useAMS]{mnras}

\usepackage{newtxtext,newtxmath}

\usepackage[T1]{fontenc}

\DeclareRobustCommand{\VAN}[3]{#2}
\let\VANthebibliography\thebibliography
\def\thebibliography{\DeclareRobustCommand{\VAN}[3]{##3}\VANthebibliography}

\usepackage{graphicx}	
\usepackage{amsmath}	
\usepackage{lscape}
\usepackage[table]{xcolor}
\usepackage[graphicx]{realboxes}
\usepackage{rotating}
\usepackage{adjustbox}
\usepackage{natbib}
\usepackage{soul}

\usepackage{hyperref}
\hypersetup{colorlinks=true, citecolor=blue, linkcolor= magenta}
\usepackage{multirow}
\usepackage{makecell}
\usepackage{natbib}
\bibpunct{(}{)}{;}{a}{}{,}
\def\apj{The Astrophysical Journal}%
\def\apjl{The Astrophysical Journal, Letters}%
\def\apjs{ApJS}%
\def\aap{Astronomy \& Astrophysics}%
\def\aaps{Astronomy \& Astrophysics, Supplement}%
\def\mnras{MNRAS}%
\def\na{New A}%
\def\pasj{PASJ}%
\def\nat{Nature}%
\def\physrep{Phys.~Rep.}%
\def\aql{Aql~X-1}
\def\sax{SAX~J1748.9-2021}
\def\hete{HETE~J1900.1-2455}
\def\maxi{MAXI~J0911-655}

\usepackage[nolist,printonlyused]{acronym}

\begin{acronym}[TDMA]
 \acro{LMXB}{Low Mass X-ray Binary}
 \acrodefplural{LMXB}{low mass X-ray binaries}
 \acro{LMXBT}{low mass X-ray binary transient}
 \acro{AMXP}{accreting millisecond X-ray pulsar}
 \acrodefplural{AMXP}{accreting millisecond X-ray pulsars}
 \acro{PCA}{Proportional Counter Array}
 \acro{ASM}{All Sky Monitor}
 \acro{RXTE}{Rossi X-ray Timing Explorer}
 \acro{MAXI}{monitor of all sky X-ray image}
 \acro{HMXB}{high mass X-ray binary}
 \acrodefplural{HMXB}{High mass X-ray binaries}
 \acro{Swift/XRT}{the Swift gamma-ray burst mission/X-ray telescope}
 \acro{Swift}{the Swift gamma-ray burst mission}
 \acro{XSPEC}{an X-ray spectral fitting package}
 \acro{FRED}{fast-rise-exponential-decay}
 \acro{DIM}{the disk instability model}
 \acro{LIS}{low-intensity-state}
 \acro{SXT}{soft X-ray transient}
 \acro{HS}{high/soft state}
 \acro{LH}{low/hard state}
 \acro{PCU}{proportional counter unit}
 \acro{ISS}{international space station}
 \acro{MSP}{millisecond pulsar}
 \acro{NS}{neutron star}
 \acro{BH}{black hole}
 \acro{Insight-HXMT}{Insight -- hard X-ray modulation telescope}
 \acro{HXMTDAS}{Insight-HXMT data analysis software}
 \acro{NXP}{nuclear powered X-ray pulsar}
\end{acronym}

\title[Spectral Study of \aql~during \textit{Pulse-on} and \textit{Pulse-off} Stages]{A Detailed Spectral Study of Intermittent-Accreting Millisecond X-ray Pulsar \aql~during  \textit{Pulse-on} and \textit{Pulse-off} Stages}

\author[T. Kocab{\i}y{\i}k et al.]{Tu\u{g}\c{c}e Kocab{\i}y{\i}k$^{1}$, 
Can G\"ung\"or$^{2,3}$\thanks{Corresponding author \newline E-mail: \href{mailto:gungor.can@istanbul.edu.tr}{gungor.can@istanbul.edu.tr} (Can Güngör)},
M. Turan Sa\u{g}lam$^{1}$,
Tolga G\"uver$^{2,3}$, \&
Z. Funda Bostanc{\i}$^{2,3}$
\\
$^{1}$\.{I}stanbul University, Graduate School of Sciences, Department of Astronomy and Space Sciences, Beyaz{\i}t, 34119, \.{I}stanbul, Turkey\\
$^{2}$\.{I}stanbul University, Science Faculty, Department of Astronomy and Space Sciences, Beyaz{\i}t, 34119, \.{I}stanbul, Turkey\\
$^{3}$\.{I}stanbul University Observatory Research and Application Center, Beyaz{\i}t, 34119, \.{I}stanbul, Turkey
}

\date{Accepted 2025 January 5. Received 2024 December 17; in original form 2024 October 16}

\pubyear{2025}

\begin{document}
\label{firstpage}
\pagerange{\pageref{firstpage}--\pageref{lastpage}}
\maketitle

\begin{abstract}
We present a detailed spectral study of an intermittent-AMXP \aql~during the \textit{pulse-on} and \textit{pulse-off} stages by using the 
archival RXTE data. 
We first perform temporal analysis by using Z$_n^2$ technique in three different energy bands, 
3.0 -- 13.0~keV, 13.0 -- 23.0~keV and 23.0 -- 33.0~keV, for the 
last 128~s time segment of the RXTE data including \textit{pulse-on} region.
We show that the pulse is the most significant in the softest band.
We, then, show that the spectrum is represented the best via combination of absorbed blackbody, disk blackbody and a gaussian line.
We modeled the last four segments of the data 30188-03-05-00 to better compare \textit{pulse-on} and \textit{pulse-off} stages.
We found a vague residual in the spectral fit of the \textit{pulse-on} segment 
between $\sim$3.0 -- 13.0~keV which agrees with the result of temporal analysis.
We show that the residual may be represented with an extra blackbody component with the temperature of 1.75~keV 
and the radius of 0.75$\pm$0.49 km.
For deeper analysis, we performed phase-resolved spectroscopy to the last 128~s, \textit{pulse-on}, segment.
We obtain two separate spectra for the spin phase range of 0.75 -- 0.25, \textit{pulse-high} and 
0.25 -- 0.75, \textit{pulse-low} and followed the same procedure.
We display that the residual becomes more clear for \textit{pulse-high} compared to the \textit{pulse-low}.
We report that the additional blackbody component, which models the residual, indicates a hotspot 
from the surface of the neutron star with the radius of 1.65$\pm$0.74 km whose temperature is 1.65 keV.
\end{abstract}
 
\section{Introduction}

The predominant X-ray sources within the Milky Way Galaxy are \acp{LMXB} comprising a compact object --a black hole or a 
neutron star (NS)-- nearly centrally located within the system, with a low-mass main sequence star 
($M_{\mathrm c}\lesssim 1\,M_{\odot}$). These systems represent the older, accreting population of the Galaxy and often 
exhibit extended accretion episodes \citep{BahDeg22}. \acp{LMXB} are further classified into various subclasses based on 
their X-ray properties, such as variability or luminosity, the characteristics of their donor stars, and the nature of the 
compact object. Mass transfer in \acp{LMXB} primarily occurs through Roche lobe overflow, where the low-mass companion star 
fills its Roche lobe, leading to material transfer from the first Lagrange point to the Roche lobe of the compact object. Due 
to the angular momentum of the transferring material and the morphology of the system, it forms an accretion disk around the 
compact object \citep{pri72}.

When a NS is born in a supernova explosion, it possesses the highest rotational velocity of its lifetime. However, 
it loses angular momentum and decelerates. Yet, within \acp{LMXB}, the mechanism of mass transfer also transfers angular 
momentum to the NS, thereby boosting its rotational speed. Consequently, \acp{LMXB} is believed 
to be the environment where millisecond pulsars are generated.
This process is known as the recycling scenario \citep{alp+82, bhat+91}.

Due to the lack of coherent pulsations, the rotation periods of NSs in almost all LMXB systems cannot be 
monitored \citep{Vaug94, Dib04, Mes15}.
The explanation of the absence of pulsations is summarized as follows: \textit{(i)} The magnetic field of the compact object 
might be too weak to canalize the material in the inner radius of the disk through the poles of the NS. \textit{(ii)} The 
Gravitational lensing effect could blur the illumination from the polar caps, then the amplitude of the signals could 
decrease to the undetectable level. \textit{(iii)} The periodic changes in the flux could be annihilated as a result of that 
X-ray light being scattered from the comptonized coronae.

Two groups differ from other \acp{LMXB} with the appearance of X-ray pulsations. 
The first is \acp{NXP} which exhibits quasi-periodic burst oscillations during the observed thermonuclear X-ray bursts.
The second is \acp{AMXP} which emits coherent X-ray pulses, with a period of milliseconds, originating from infalling plasma 
to the magnetic poles of the rotating NS from the inner layers of the accretion disk.
Only 23 \acp{AMXP} have been detected so far \citep{StrohKeek17, CampSal18, PatWat21, SalvoSanna20,SannaBult2022, Bult22}. 
The \ac{AMXP} family has a subclass, so-called intermittent-\acp{AMXP}, which exhibit discontinuous pulse phenomena 
with \textit{pulse-on} and \textit{pulse-off} stages. They are unique 
labs to comprehend the absence of the pulse behavior in \ac{LMXB} systems. This subclass has only four members 
yet, \hete, \sax, \aql, and \maxi, respective to their discovery.

\aql~is a NS-LMXB whose donor is a K4±2 main sequence star that rotates with a period of 18.95~hr \citep{ChevIl91} in a 
36$^\circ$ -- 47$^\circ$ inclined orbit \citep{mata17}.
In the light of \ac{RXTE} data, it has been reported that the distance of the source is \hbox{4.4 -- 5.9} kpc derived from 
type-I photospheric radius 
expansion bursts assuming the Eddington limit is 2.0 -- 3.8 $\times 10^{38}$~erg~s$^{-1}$ \citep{jon04}. 
It is also classified as a \ac{SXT} due to its almost annual cyclic outbursts with a recurrence time of 60 to 140 days and a 
duration of 25 to 60 days \citep{sim02, gungor+17a}.
In the quiescent state, the X-ray Luminosity is $L_{\rm   X}  \approx   10^{33}$~erg~s$^{-1}$  \citep{verb+94} 
while the outburst peak luminosity could reach up to $L_{\rm X} \approx 10^{37}$~erg~s$^{-1}$ \citep{cam+13}. 
\aql~is a unique member of the Intermittent-\acp{AMXP} \citep{koy+81} with 550.27~Hz pulse frequency, consistent 
with the burst oscillations \citep{cas+08}, which has been detected only for 150~s overall 20 years in the peak of 1998 
outburst.

In this work, to enlarge our knowledge about pulse phenomena, 
we applied simultaneous timing and spectral analysis to the pulsation episode of \aql. 
We explain the detailed data reduction, timing, and spectral analysis using the \ac{RXTE} data covering the pulse episode, in 
\autoref{obs}. We discuss and conclude our present results of the timing and spectral analysis in \autoref{discuss-conc}.


\section{Data Analysis and Results}
\label{obs}

We used the data of \ac{RXTE}/\ac{PCA} detector \citep{Jahoda2006}, operating within its most 
sensitive energy range of 3.0 -- 30.0~keV. The PCA \citep[2.0 -- 60.0~keV;][]{Bra1993} is particularly suited for studying 
spectral variation on short timescales, such as pulsations or X-ray Bursts, thanks to its large collecting area ($\simeq$6500 
cm$^2$) and high time resolution down to $\mu$s. 
Our analysis involved systematic timing and spectrum evaluations focused on the pulsed episode of \aql.
Additionally, we examined the spectral evolution before, during and after the pulsation.
Essential details regarding the data used in this study are 
summarized in \autoref{tabx}.

\begin{table}
\caption{Elementary information of the RXTE/PCA data of the sample set in this study.}
\setlength{\tabcolsep}{0.75\tabcolsep}
  \centering
\begin{center}
\begin{small}
\begin{tabular}{ccccc}
\hline
\textbf{Obs} &\textbf{ObsID/SE$^\ast$ }& \textbf{Exposure}	& \textbf{Start Time} & \textbf{Count Rate}  \\
&	& \textbf{(ks)}	& \textbf{(MJD)}& \textbf{(cnt/s)} \\
\hline
ObsID1&30188-03-04-00/1 &	1.294 & 50881.51507897 &  $3748.83$ \\							
ObsID2&30188-03-05-00/2 &	1.596 & 50882.93530433 &  $4296.37$ \\							
ObsID3&30071-01-01-01/1 & 	1.832 & 50884.04936646 &  $ 3959.26 $ \\						
\hline
\multicolumn{4}{l}{$^\ast$ Science Event Id.}
\end{tabular}
\end{small}
\label{tabx}\\
\end{center}
\end{table}

\begin{figure}   
\centering
  \includegraphics[angle=0,width=0.47\textwidth]{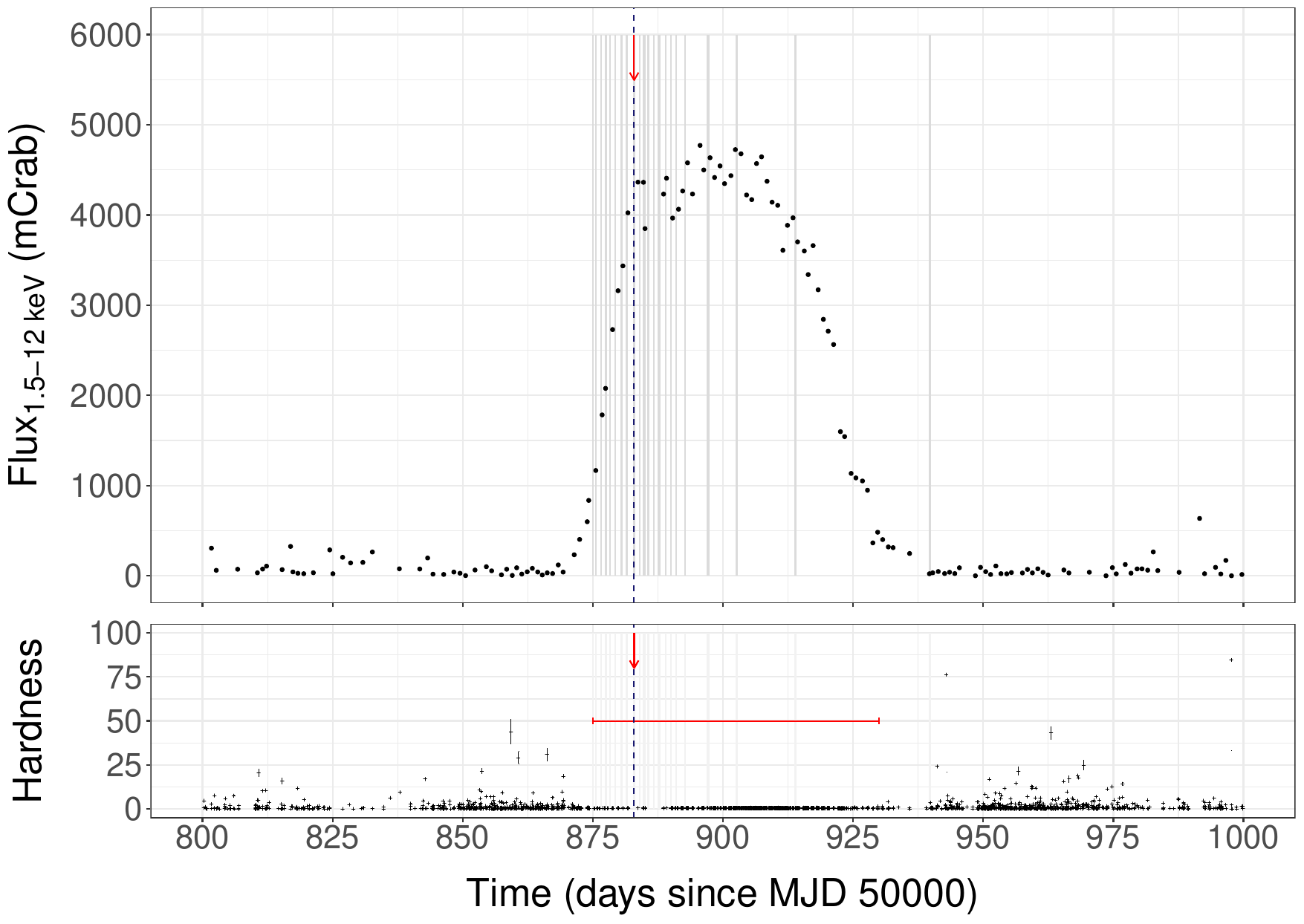}
     \caption{The light curve of the outburst of \aql~in 1998 via ASM data (upper panel, black points) with the time evolution of the hardness ratio
     estimated by using the ratio of count rates in the energy ranges of 5.0 -- 12.0~keV and 1.5 -- 5.0~keV (bottom panel, black points).
     The times of the pointing RXTE observations and the detected pulsation are shown via vertical grey lines and the blue dashed line, respectively.}
         \label{lc_hardness}
\end{figure}

The All-Sky Monitor \citep[ASM;][]{asm95} light curve of the 1998 outburst of \aql, whose duration is about eight months, is 
displayed in \autoref{lc_hardness}. The hard count rates are derived from ASM C band data in the energy range of 5.0 -- 12.0~keV, while 
the soft count rates are computed using the sum of the data of ASM A and B bands in the energy range 
of 1.5 -- 3.0~keV and 3.0 -- 5.0~keV, respectively. The lower panel of \autoref{lc_hardness} illustrates the evolution of the 
hardness ratio. Notably, the range delimited by the red line on the hardness curve indicates 
the soft-high state during the outburst.
This state signifies accretion dominance, with the observed pulse phenomena within this regime.

These three consecutive data given in \autoref{tabx} are split into 128~s time windows
for both temporal and spectral analysis. One of the reasons for this segmentation strategy is to avoid
computational obstacles by choosing a binary number to create spectra which result to get response and background files in the 
same length with extra effort. Beside this, although the pulse duration is reported as slightly longer than 150~s,
the power spectrum given in \citet{cas+08} is also for 128~s long though our 
results would be comparable with the ones in the literature in which our search is performed via Z$_n^2$ technique while
the mentioned one in the literature is via classical Fourier method.

\subsection{Temporal Analysis}
\label{timin}

We began with employing the \textit{faxbary} 
task in \textit{ftools}\footnote{A General Package of Software to 
Manipulate FITS Files; \hyperlink{https://heasarc.gsfc.nasa.gov/ftools/}{https://heasarc.gsfc.nasa.gov/ftools/}} 
package to apply the barycentering correction to photon arrival times, before performing temporal analysis to the
observations listed in \autoref{tabx}, 

\begin{figure*}
\centering
  \includegraphics[angle=0,width=0.75\textwidth]{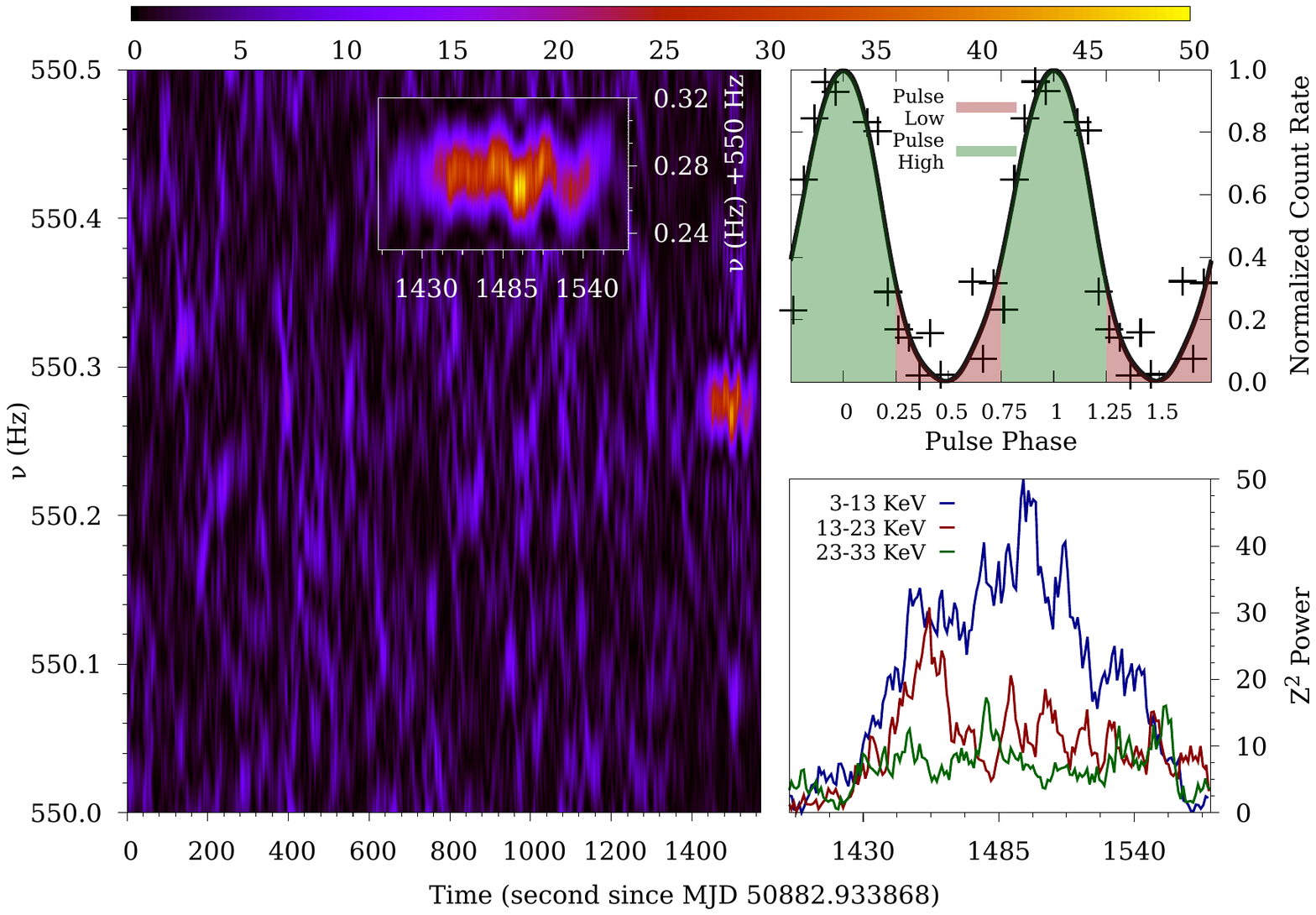}
  \caption{The dynamic power spectrum of ObsID 30188-03-05-00 of \aql~with a zoomed-in view to the last 150~s (pulse-on) in the inset (The left panel)
  in the energy range of 3.0 -- 13.0~keV. Pulse profile for the last 150~s with normalized count rates is presented in the upper right panel while the defined \textit{pulse-low} and \textit{pulse-high} are shown as shaded pink and green areas.
  The time evolutions of Z$_1^{2}$ power values at 550.27~Hz during the last 150~s in the energy ranges of \hbox{3.0 -- 13.0~keV} (blue), 13.0 -- 23.0~keV (red), and 23.0 -- 33.0~keV (green) are given in the lower right panel.
  }
\label{powerspectra}
\end{figure*}

For the first step, we utilized the Z$^{2}_{n}$ statistic \citep{Buc83}. This approach is preferred over alternatives because it operates directly on photon arrival times rather than on binned data. The power of Z$^{2}_{n}$ statistic is defined as;
\label{temp}
\begin{equation}
Z^2_n=\frac{2}{N} \sum_{\rm k=1}^n\left[\left(\sum_{\rm j=1}^N \cos k \Phi_{\rm j}\right)^2+\left(\sum_{\rm j=1}^N \sin k \Phi_{\rm j}\right)^2\right]
\label{equ:z2}
\end{equation}
where Z$^{2}$ represents the power, N denotes the total number of photons in the search interval, n stands for the number of 
harmonics (here, it is set to be 1), and $\Phi$ represents the phase of the photons.

Subsequently, we computed Z$_1^{2}$ power spectra within the 3.0 -- 30.0~keV energy band for 
each segment of the three observation sets, each lasting 128~s. 
The $Z_1^2$ technique can perform a direct temporal analysis on the arrival time of the photons and allows us to reach 
the time resolution limit of the detector, 1~$\mu$s for PCA, as long as there are enough photons in the sample set.
We used the previously reported frequency of $550.27~\text{Hz}$ for the source and performed the scan
in $10^{-4}~\text{Hz}$ steps in the frequency range of $550.0$ -- $550.5~\text{Hz}$.
This frequency resolution is sufficient to accurately determine the detected pulse frequency.

Our analysis identified a significant power at 550.27~Hz only in the last segment of ObsID 30188-03-05-00, consistent with 
the report in \citet{cas+08}. 
The maximum Z$_1^{2}$ power of the signal is 104.79, with a single trial probability of 1.7 $\times 10^{-23}$ calculated 
by assuming a Poisson noise distributed as $\chi^2$ with two degrees of freedom. 
Considering the number of trials, the number of frequency bins multiplied by the number of power spectra, 
we determined the chance occurrence probability of the signal at 550.27~Hz to be 3.6$\times 10^{-18}$.

Same as in the literature, we did not detect any sign of possible pulse in the power spectra generated for the previous and 
latter data of the pulse detected ObsID. For a more detailed analysis of the pulsation, we computed a dynamic power spectrum 
using a search time window of 25~s, shifting by 1 s along the entire observation. We again performed our search in the 
frequency range of 550.0 -- 550.5~Hz with a resolution of 10$^{-4}$ Hz.
Since the most spectrally effective energy range of PCA is 3.0-30.0 keV, we performed all temporal analysis
starting from 3.0~keV to be consistent with the further spectral analysis.
Besides this, sub-setting the data for 25~s time windows decreases the photon count to make good Z$_1^2$ statistic.
Therefore, as a semi-arbitrary optimal choice, we divided the energy range into equal three parts,
3.0 -- 13.0~keV, 13.0 -- 23.0~keV, and 23.0 -- 33.0~keV,
to get an idea about the energy range that the pulse is the most significant.
We track the Z$_1^2$ power values instead of pulse fractions which might
be misleading since the count rates are very low for higher energies.
The pulse with the maximum power was observed in the energy range of 3.0 -- 13.0~keV. The 
resulting dynamic spectrum in 3.0 -- 13.0~keV is presented in the left panel of \autoref{powerspectra}. It is noticed that 
the power has increased significantly in the last 150~s, as depicted in the inset. We also displayed the pulse profile 
derived by folding the light curve with the 550.27~Hz frequency in the upper right panel of the figure. The green region 
indicates \textit{pulse-high}, while the pink denotes \textit{pulse-low} which will be used to perform our further 
``phase-resolved spectroscopy'' discussed later in \autoref{spec-rxte}. The lower right panel illustrates the time evolution of 
pulse power in the three energy ranges.
We would like readers to pay attention to that the power is substantially higher in the 3.0 -- 13.0~keV energy band compared 
to the other bands.

\subsection{Spectral Analysis}
\label{spec-rxte}

We utilized the "saextrct" task in ftools to extract clean light curves and spectra.
We independently generated response matrices for the three datasets provided in \autoref{tabx} using PCARSP version 11.7.  
HEASOFT version 6.30.1 is used to analyze all data sets.
We generated background spectra using the module file for bright sources, pca-bkgd-cmbrightvle-eMv20051128.mdl.
Spectra were obtained by classifying all channels of at least 20 counts.
To take into account instrumental uncertainties, we added a systematic error of 1.0\%.
We applied abundances from \citet{wilm2000} to take into account the 
interstellar medium, interstellar grains, and the H$_{2}$ molecule. In XSPEC version 12.0, we used \textit{phabs} as a 
photoelectric absorption model with the default cross-section \citep{ver+96}.
The neutral hydrogen column densities were fixed to $N_{\rm H}~=~3.4\times 10^{21}~{\rm cm}^{-2}$ 
following \citet{mac03spec}.
We employed the distance of 5 kpc to transform normalization values to radii \citep{jon04}.


 \subsubsection{Step 1: Model Selection}
 \label{step1}

Unlike non-pulsating LMXBs, which typically show transitions between hard and soft spectral states,
AMXPs are often characterized as hard X-ray transients. 
The X-ray continuum generally consists of one or two blackbody-like components and an unsaturated Comptonization component 
\citep{GierDone02, GierPou+05, Pout+06}. In the hard state, the spectrum is dominated by a hard/Comptonized component along 
with a soft/thermal component, whereas in the soft state, the spectrum is primarily dominated by the soft/thermal component 
\citep{lin+07}.

One of the main objectives of this study is to understand the spectral effect of the origin of the pulsation.
The most important step is to model the \textit{pulse-off} spectrum with very high accuracy,
possibly even more than what model is used, to get the discrepancy with the \textit{pulse-on}.
To have a sense of which model would best describe the data, 
we first tracked hardness evolution and established the source's spectral state.
Given the time evolution of hardness (\autoref{lc_hardness}), we considered three models commonly used to describe the 
source in the soft state: phabs$^\ast$(diskbb+nthcomp+gau) \citep{rai+11, sak+12, abdel+21}, phabs$^\ast$(diskbb+comptt+gau) 
\citep{lin+07, gogus07}, and phabs$^\ast$(diskbb+bbodyrad+gau) \citep{Gung20}.
We then applied these models to each 128~s segment of the data listed in \autoref{tabx}. 

We started with a fundamental model of \textit{bbodyrad+diskbb} to explain a simple accretion scenario 
and noticed the model efficiently works as previously mentioned in the literature \citep{Gung20}.
Previous studies \citep{mitsu89, GierDone02, lin+07} have shown that residuals above 15.0~keV indicate 
reverse Compton effects to take into account up-scattering photons by possible corona.
We, though, proceeded a combination of a comptonization model \citep[\textit{compTT} in XSPEC,][]{comptt}
with a disk blackbody in which the Wien temperature is linked to the temperature of the disk blackbody model 
under the assumption of that photons from the inner accretion disk are seeds for the inverse Compton process.
The plasma temperature of \textit{compTT} is fixed to a reasonable value of 15.0~keV \citep{lin+07}.
The geometry is chosen as a disk \citep[see][for details]{titar95}.
We also attempted \textit{nthcomp} \citep{zdziar96, zyck99} as a Comptonisation model 
which allows us to switch blackbody and disk blackbody as the seed photons 
and to investigate relatively larger range of optical depths.
Following the studies in the litarature with similar strategy \citep{rai+11, sak+12, abdel+21},
we applied the model as \textit{diskbb}+\textit{nthcomp}$_{\text{bb}}$,
which approves the two optically thick components, T$_{\text{in}}$ and T$_{\text{bb}}$.
This implies the situation of where the blackbody emission from the NS surface is fully comptonized,
while the disk emission is observable.
The electron temperature is again fixed to 15.0~keV while T$_{\text{in}}$, T$_{\text{bb}}$ and $\Gamma$ are kept as free fit parameters.

After constructing all of the models, we identified the most efficient model with best $\chi^{2}$ and reasonable physical parameters such as temperature, normalization etc.
Among these, we identified the \textit{phabs$^\ast$(diskbb+bbodyrad+gau)} model as the most suitable 
and as simple as possible to well fit the data.
In this model, the {\it bbodyrad} component represents the boundary layer emission from the NS surface, {\it diskbb} 
represents the emission from the accretion disk, and {\it gau} represents the broad iron line originating from the accretion 
disk. As noted in \citet{lin+07} and \citet{Asai2000}, the line energy can be constrained to 6.2 -- 7.3 keV with $\sigma$ 
ranging from 0.1 to 1.0 for the Gaussian model. To reduce the number of free parameters in our analysis, we fixed the line 
energy at 6.4 keV with a $\sigma$ of 0.8 for all segments. 
To test the hypothesis that a Compton Cloud could be responsible for the absence of the pulse, we added a 
\textit{comptt}  model \citep{comptt} to our chosen model and examined the variation in the parameters.
Comparing the model combinations, we found no significant differences in the evolution of the model parameters statistically.
The results suggest that high accretion may disperse the Compton cloud, supporting the conclusion that the 
phabs$^\ast$(diskbb+bbodyrad+gau) model is the most reliable for our study. Therefore, we conclude that the lack of a 
detectable pulse is not due to scattering by the Comptonized corona, and this scenario does not directly explain the pulse
phenomena.

\subsubsection{Step 2: Segmented Sequential Spectral Analysis}
\label{step2}

\begin{figure}
\centering
\includegraphics[angle=0,width=0.5\textwidth]{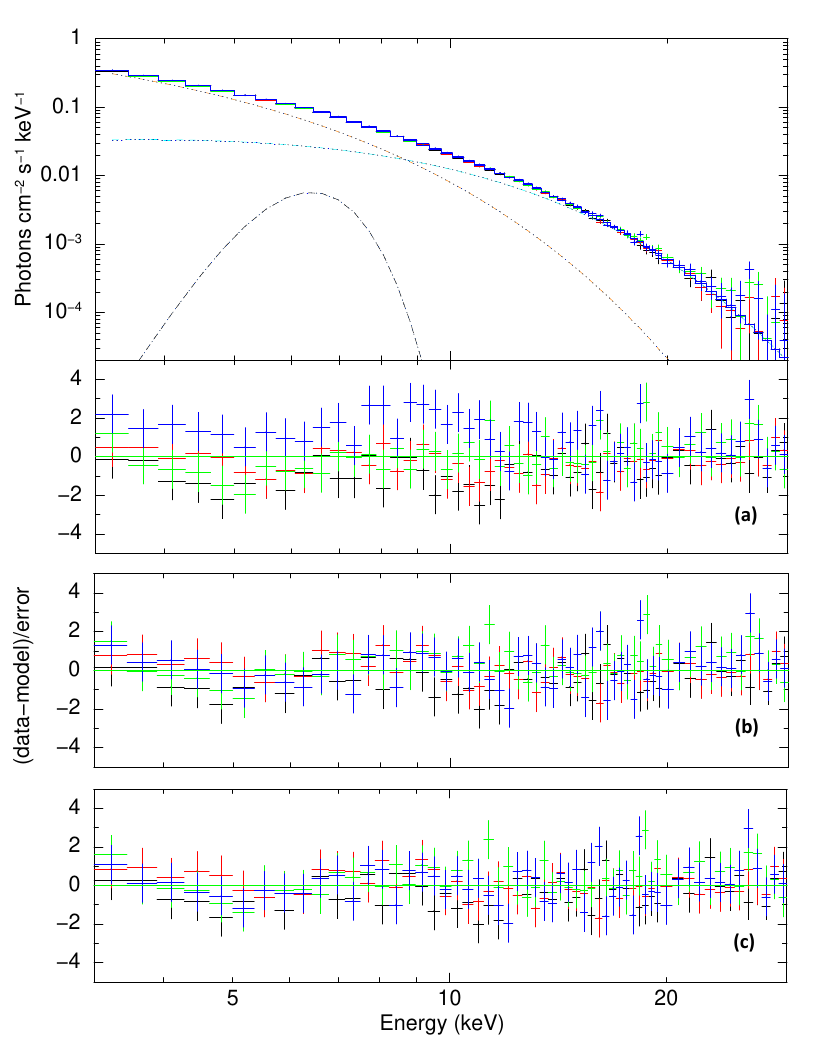}
     \caption{The unfolded X-ray spectrum of \aql~for last four segments, 9 (black), 10 (red), 11 (green), 12 (blue), with the components from the best model (upper panel) and residuals for three approaches; 
     (a) The four segments are modelled as linked to each other (the second panel), 
     (b) The pulse-on segment is fit independently from the previous three segments (the third panel), 
     (c) The last segment is linked to previous three segments with an additional independent blackbody component (the bottom panel).}
         \label{128_spec}
\end{figure}

Sources like \aql~exhibit rapidly varying spectral characteristics. To detect potential variability caused by the pulse, we performed a simultaneous fit 
for the last four segments (9$^{\text{th}}$, 10$^{\text{th}}$, 11$^{\text{th}}$ and 12$^{\text{th}}$ segments of ObsId 30188-03-05-00) of the 
data including \textit{pulse-on} region. Notably, only the final segment (12$^{\text{th}}$) shows pulsation.
The results of the simultaneous fit for these four segments, as well as the normalization values with blackbody temperature when included, 
are summarized in \autoref{tab2} and shown in \autoref{128_spec}.

Firstly, we simultaneously fit the \textit{pulse-off} segments (9$^{\text{th}}$, 10$^{\text{th}}$ and 11$^{\text{th}}$) and then we used the results
as input fit parameters to the \textit{pulse-on} segment (12$^{\text{th}}$).
As can be seen in panel (a) of \autoref{128_spec}, the residuals in the \textit{pulse-on} segment (blue pluses) show 
slightly higher scatter than the other segments in the softer energy.
It is seen in the first row of \autoref{tab2} that the $\chi^2$ value of this scatter is above 3.
Remembering the energy range where the pulse is the most powerful is 3.0 -- 13.0 keV, the picture in the panel (a) of \autoref{128_spec}
strengthen the idea that the spectrum needs an extra component for \textit{pulse-on} characteristic.

Secondly, apart from the first approach, we only set the blackbody and disk blackbody parameters of the \textit{pulse-on} segment free in order to 
determine any variation between the \textit{pulse-on} and \textit{pulse-off} regions.
As can be seen in the second row of \autoref{tab2}, model values of both \textit{pulse-off} and \textit{pulse-on} segments are statistically consistent 
to each other (the panel (b) of \autoref{128_spec}). 
We, then, refitted the segments by adding 
an extra blackbody component only on \textit{pulse-on} one, assuming this excess is caused by the area where the pulse is generated.
We set the temperature and normalization of the additional `bbodyrad' component  unfrozen for the \textit{pulse-on} segment.
We obtained the temperature to be $1.75\pm{0.20}$ keV from the additional blackbody.
In the light of this approach, which can be seen in the last row of \autoref{tab2} and the panel (c) of \autoref{128_spec},
the parameters were improved physically and statistically to make them consistent. 
It is interpreted that the NS boundary layer provided the blackbody component, the pulse provided the secondary blackbody,
and the accretion disk provided the disk blackbody. 

\begin{figure}
\centering
  \includegraphics[angle=0,width=0.505\textwidth]{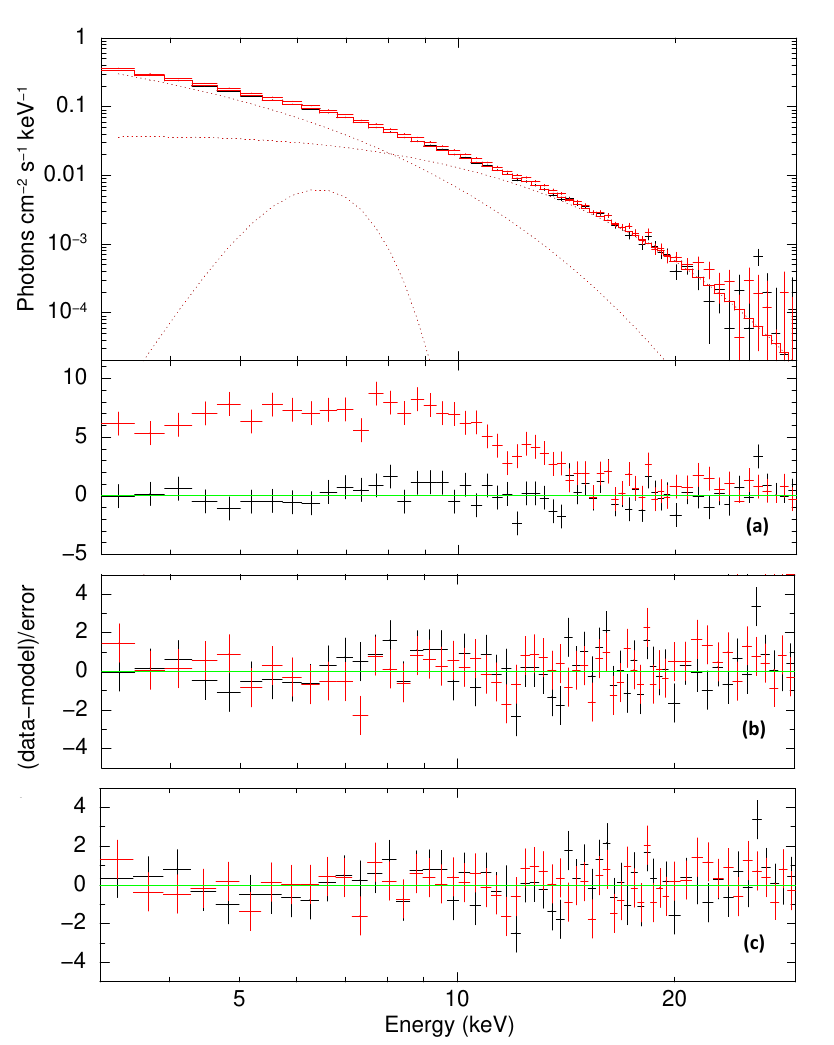}
     \caption{The unfolded X-ray spectrum of \aql~for \textit{pulse-low} (black) and \textit{pulse-high} (red) spectra, with the components from the best model (upper panel) and residuals for three approaches; 
     (a) The \textit{pulse-low} and the \textit{pulse-high} spectra are modelled as linked to each other (the second panel), 
     (b) The \textit{pulse-high} spectrum is fit independently from the \textit{pulse-low} one (the third panel), 
     (c) The \textit{pulse-high} is linked to the \textit{pulse-low} with an additional independent blackbody component (the bottom panel).}
         \label{phase_spec}
\end{figure}

\begin{table*}
\caption{The best simultaneous fit parameters of the model of \textit{blackbody} \textit{(}+ \textit{blackbody)} + \textit{disk blackbody} + \textit{Gauss} for the last four 128~s segments.}
\setlength{\tabcolsep}{2.0\tabcolsep}
\centering
\begin{center}
\begin{scriptsize}
\begin{tabular}{ccccccccc}

\textbf{Approach} &\textbf{SegID}	&	\textbf{kT$_{bb}$}	&	\textbf{Norm$_{bb}$} &	\textbf{kT$_{dbb}$}	& \textbf{Norm$_{dbb}$} & \textbf{kT$_{2bb}$}	& \textbf{Norm$_{2bb}$} &	\textbf{$\chi^2$/dof} \\
\hline
First$^\text{a}$ & {9/10/11/12$^\dagger$} &$	2.24	\pm{	0.06	}$ & $	10.27	\pm{	2.20	}$ & $	1.53	\pm{	0.05	}$ & $	98.59	\pm{	12.89	}$ & $	- $ & $ - $ &  0.69/0.38/0.83/3.04	\\
\hline
Second$^\text{b}$ &{9/10/11}	&$	2.24	\pm{	0.06	}$ & $	10.27	\pm{	2.20	}$ & $	1.53	\pm{	0.05	}$ & $	98.59	\pm{	12.89	}$  & $ - $ & $ - $ &  0.69/0.38/0.83 \\	
&{12$^\dagger$}	&$	2.26	\pm{	0.13	}$ & $	9.77	\pm{	4.21	}$ & $	1.57	\pm{	0.09	}$ & $	90.82	\pm{	22.36	}$  & $ - $ & $ - $& 0.96\\	
\hline
Third$^\text{c}$ &{9/10/11}	&$	2.24	\pm{	0.06	}$ & $	10.27	\pm{	2.20	}$ & $	1.53	\pm{	0.05	}$ & $	98.59	\pm{	12.89	}$ & $ - $ & $ - $ & 0.79/0.38/0.83	\\
&{12$^\dagger$}	&$	2.24	\pm{	0.06	}$ & $	10.27	\pm{	2.20	}$ & $	1.53	\pm{	0.05	}$ & $	98.59	\pm{	12.89	}$ & $	1.75	\pm{	0.20	}$ & $	2.24	\pm{	0.96	}$ &	0.97	\\
\hline
\end{tabular}
\footnotesize{\begin{flushleft}$^\text{a}$Four segments are linked. $^\text{b}$The last segment is independent of the previous three segments. $^\text{c}$The last segment is linked to previous three segments with an additional independent blackbody component.\\
$^\dagger$The \textit{Pulse-on} segment\end{flushleft}}
\end{scriptsize}
\label{tab2}
\end{center}
\end{table*}

\begin{table*}
\caption{The best fit parameters of the model of \textit{blackbody} \textit{(}+ \textit{blackbody)} + \textit{disk blackbody} + \textit{Gauss}
for phase-resolved spectroscopy.}
\setlength{\tabcolsep}{2.0\tabcolsep}
  \centering
\begin{center}
\begin{scriptsize}
\begin{tabular}{ccccccccccc}
\textbf{Approach} &\textbf{SegID}	&	\textbf{kT$_{bb}$}	&	\textbf{Norm$_{bb}$} &	\textbf{kT$_{dbb}$}	& \textbf{Norm$_{dbb}$} & \textbf{kT$_{2bb}$}	& \textbf{Norm$_{2bb}$} &	\textbf{$\chi^2$/dof} \\
\hline
First$^\text{a}$ &12$_{low}$/12$_{high}$	&$	2.19	\pm{	0.13	}$ & $	11.98	\pm{	4.87	}$ & $	1.48	\pm{	0.11	}$ & $	109.84	\pm{	33.45	}$ & $-$ & $-$ &	0.98/20.25	\\							
\hline
Second$^\text{b}$ &12$_{low}$	&$	2.19	\pm{	0.13	}$ & $	11.98	\pm{	4.87	}$ & $	1.48	\pm{	0.11	}$ & $	109.84	\pm{	33.44	}$ & $-$ & $-$ & 	1.07	\\
&12$_{high}$ &$	2.28	\pm{	0.22	}$ & $	9.23	\pm{	6.31	}$ & $	1.62	\pm{	0.13	}$ & $	83.38	\pm{	28.21	}$ & $-$ & $-$ & 	0.74	\\	
\hline
Third$^\text{c}$ &12$_{low}$	&$	2.19	\pm{	0.13	}$ & $	11.98	\pm{	4.87	}$ & $	1.48	\pm{	0.11	}$ & $	109.84	\pm{	33.44	}$ & $ - $ & $ - $ & 0.98	\\
&12$_{high}$ 	&$	2.19	\pm{	0.13	}$ & $	11.98	\pm{	4.87	}$ & $	1.48	\pm{	0.11	}$ & $	109.84	\pm{	33.45	}$ & $	1.65	\pm{	0.06	}$ & $	10.88	\pm{	2.18	}$ & 0.84 \\
\hline

\end{tabular}
\footnotesize{\begin{flushleft}$^\text{a}$\textit{The pulse-low} and \textit{the pulse-high} spectra are linked. $^\text{b}$The pulse-high spectra is independent of the \textit{pulse-low} one. $^\text{c}$The \textit{pulse-high} spectra is linked to the \textit{pulse-low} spectra with an independent blackbody component.\end{flushleft}}
\end{scriptsize}
\label{tab3}
\end{center}
\end{table*}
\subsubsection{Step 3: Phase-Resolved Spectroscopy}
\label{step3}

To put forward more clear picture about the effect of the \textit{pulse-on} to the spectrum, we performed phase-resolved 
spectroscopy within millisecond resolutions.
We used only the last 128~s within the last 150~s where the pulse is seen, neglecting 10~s from the end. 
Since the detection of the pulse takes place at the end of the observation, neglecting the last 10~s minimizes the 
amount of possible errors that may arise from instrumental effects.
With this preference, we also eliminated the transitional effects at the beginning and terminal.
Though, the time segment of the pulse with higher and relatively constant amplitude is used for phase-resolved spectroscopy.
We created the pulse profile as an output of our temporal analysis. The phase range of 0.25 -- 0.75 and 0.75 -- 0.25 are chosen 
as \textit{pulse-high} and \textit{pulse-low} regions as can be seen as green and pink colored areas in the right-upper panel 
of \autoref{powerspectra}, respectively. We then created two independent spectra for both \textit{pulse-high} 
and \textit{pulse-low} from last 128~s segment, so each has 64~s duration.

Following the similar procedure as in \autoref{step2}, we first fitted the \textit{pulse-low} (the \textit{pulse-off} in the case of \autoref{128_spec}) 
spectra by using the model of \textit{phabs*(bbodyrad+diskbbody+gau)}. Then, we used the best-fit parameters as frozen inputs adding an extra blackbody component to 
fit the \textit{pulse-high} (the \textit{pulse-on} in the case of \autoref{128_spec}). Therefore, the panel (a) of \autoref{phase_spec} shows the residuals 
without the extra blackbody, comparing the pulse-high spectrum to the best model of pulse-low without free normalization since these are intertwined data.
This strategy allows us to determine the residuals with greater precision.
Also, as it can be seen in the first approximation of \autoref{tab3}, the value of $\chi^2$ is determined as 20.35 which indicates much higher residuals.
We set the blackbody and disk blackbody parameters only of the \textit{pulse-high} free in order to specify any differences 
between the \textit{pulse-high} and \textit{pulse-low} spectra. 
The results of the free parameter deviation for both segments are given in the \autoref{tab3} as the second approach.
The model parameters for both \textit{pulse-low} and \textit{pulse-high} segments are statistically consistent with each other,
as can be seen in the residuals shown in panel (b) of \autoref{phase_spec}.
We finally refitted only the \textit{pulse-high} segment by adding an extra blackbody component 
assuming that the extra radiation comes from the hot spot which is the cause of the observed coherent pulsation.
The temperature of the extra blackbody component is obtained to be about ~$1.65\pm{0.06}$ keV whose radius of ~$1.65\pm0.54$ km indicated by its normalization.
Based on this approach, as presented in the last row of \autoref{tab3} and panel (c) of \autoref{phase_spec},
the parameters were improved both physically and statistically comparing to the second approach.

\section{Discussion and Conclusions}
\label{discuss-conc}
In this study, we have applied a 3-step analysis to demonstrate the existence of the spectral effects of the pulse phenomenon.
We present the output of the spectral analysis of three \aql~observations obtained by \ac{RXTE} in \autoref{step1}.
Firstly, we modelled the spectra of \aql~in the energy range of 3.0 -- 30.0~keV with the models previously used in the literature and determined the most appropriate model.

Then, for more detailed analysis, we compared the \textit{pulse-on} segment with the \textit{pulse-off} ones.
Due to the high spectral variability of sources like \aql, we modeled the last four segments simultaneously, 
in which the last is the \textit{pulse-on} and the rest is \textit{pulse-off}.
In the spectral fit of the last segment, we detected a discrepancy, which we thought as the effect of the pulse,
so we added another blackbody to the model.
By setting the added blackbody parameters free, we obtained the temperature of about 1.75~keV. In this case, the blackbody component came from the NS boundary layer,
the secondary blackbody is from the pulse while the disk blackbody component is the contribution of the inner layers of the accretion disk.
\citet{cas+08} stated that a combination of disk black body with a component representing the boundary layer 
is sufficient to model the whole spectrum of Aql X-1. 
Our spectral analysis agrees with that a combination of two physical models may well fit the data.
At this point, we prefer to use the blackbody component to express the boundary layer and the disk blackbody for the disk 
by adding a Gaussian to represent the iron line. 
We also present two model solutions in panels (b) of \autoref{128_spec} and \autoref{phase_spec}.
However, especially in phase-resolved spectroscopy, spectra for \textit{pulse-low} and \textit{pulse-high} are constructed 
from data intertwined in millisecond intervals. 
Though, any difference would not be expected in the fit parameters, even the normalization.
Under this assumption, we followed the strategy of adding an extra blackbody component and found that the  $\chi^{2}$ values well improved. 

Finally, in \autoref{step3}, we applied phase-resolved spectroscopy to the 128~s \textit{pulse-on} segment.
For this reason, using the spin frequency and ephemeris values of the source obtained from the temporal analysis,
we created \textit{pulse-high} and \textit{pulse-low} spectra for the phase ranges of 0.25 -- 0.75 and 0.75 -- 0.25, respectively.
The residual detected in the previous step becomes more clear revealed, as shown in \autoref{phase_spec}.
In agreement with the results we obtained in the previous step, we added a blackbody component for the \textit{pulse-high} and obtained 
that the pulse could have a source with a temperature of about 1.65~keV.
We utilized the \textit{bbodyrad} task in XSPEC to simulate the difference between the \textit{pulse-low} and \textit{pulse-high} 
spectra using a single additional blackbody component. The hot spot is assumed to be circular in this task.
Although only spectral results cannot clearly confine the location, the intermittency of the pulse suggests the location to be at 
very high latitudes with very low angles between the rotating axis \citep{Lamb2009a}.
 
Examining \autoref{lc_hardness}, we see that the pulsation becomes visible near the outburst peak when the source
is in the soft state. The unabsorbed flux for 3.0 -- 30.0~keV is calculated as $8.6\times 10^{-9}$~erg~s$^{-1}$~cm$^{-2}$.
This corresponds to a total amount of flux that describes the model we used, and also the contribution attributed to the 
pulsation blackbody component is calculated as $7.4 \times 10^{10}$~erg~s$^{-1}$~cm$^{-2}$, meaning it corresponds to 8.62\% of 
the total flux.
This value is also related to the pulse profile as well as the peak -or pulse fraction- of the flux.
In the meanwhile, the difference between the \textit{pulse-low} and \textit{pulse-high} spectra depends on 
the definition of the spin phase, which are 0.75-0.25 and 0.25-0.75, respectively.
We obtained the pulse fraction as $\sim$4.5\%. Although \citet{cas+08} reported that the pulse fraction varies between
$\sim$2\% and $\sim$10\%, the pulse profile given in the inner panel of Figure 1  of \citet{cas+08}
indicates a pulse fraction of $\sim$4.0-4.2 with the minimum of the pulse profile is $\sim$4750 cnt/s and 
the maximum is $\sim$4950 cnt/s.
Our estimation of the pulse fraction is consistent with this value.
Given the sinusoidal shape of the pulse profile, it may not be very easy to just relate the pulsed fraction
to the fraction of the flux of the blackbody component to the total flux. Furthermore, while the blackbody model flux
contributes to some extend to the whole energy band that RXTE is sensitive, the pulsed fraction is only calculated within a 
limited energy range we defined. In addition to these observational effects  there should be other factors that needs to be 
taken into account as well. For example our  spectral analysis do not take into account the gravitational redshift or the very 
fast rotation of the NS into account. These should also need to be taken into account if we aim to directly compare 
the flux ratio to the pulsed fraction.
 \section*{Acknowledgements}
 This work has made use of ASM data of NASA’s \ac{RXTE} satellite. This work is partially supported by the Scientific and Technological Research Council of Turkey (TUBITAK) Grant No. 120F094. CG acknowledges support from Bilim Akademisi - The Science Academy, Turkey under the BAGEP program and TK acknowledges support from TUBITAK-BIDEB 2210 fellowship.
 \section*{Data Availability}
 This study utilized observational data that is publicly available from the \ac{RXTE}/\ac{PCA} Master Catalogue of the NASA/HEASARC archive.



\bibliographystyle{mnras}

\bsp	
\label{lastpage}
\end{document}